# BeiDou-G2: Past and Present

Hou-Yuan Lin[1,2,*]

**Abstract**

In January 2022, the defunct satellite BeiDou-G2 was pulled out of geostationary orbit by Shijian-21 to a graveyard orbit. For safe docking and operation, it was necessary to determine the rotation state in advance. In this paper, we show the evolution of the rotation of the BeiDou-G2 satellite based on the photometry observation data for the past 10 years. The rotational speed of BeiDou-G2 was found to be annual oscillation, mainly due to the solar radiation. Based on the evolution of BeiDou-G2's rotation speed and its orbit, we confirmed that in the last 10 years, the satellite had six abnormal events. These abnormal events were mainly due to the increase in the rotation speed caused by suspected fuel leakages. Additionally, the abnormal events included one collision in 2012, which was inferred to be the trigger of the fuel leakages in the following years. No rotational speed abnormalities occurred again after 2017, probably due to the complete release of the residual fuel. The parameters and the propagating models after one incidence of solar panel damage in 2014 and one fragment in 2016, with the standard errors for propagating over 1 year of the rotational axis less than 3° and rotational speed being 0.11°/s, were believed to be able to satisfy the accuracy requirements of the rotation state well at the moment of docking.

The BeiDou-G2 satellite (Compass G2) is the second spacecraft of the second generation BeiDou/Compass navigation satellite system. It was launched in 2009, but it became inactive 1 year later and began to drift in an extensive range in the geosynchronous ring [1]. It was a huge threat to satellites in geosynchronous orbit (GEO), which is valuable orbital space for weather monitoring, communications, and surveillance. Hence, in January 2022, BeiDou-G2 was pulled out of the heavily populated GEO orbit range by Shijian-21 to the orbit ranging from 290 km to 3100 km above GEO in graveyard orbit. Shijian-21 was launched in 2021 for the validation of on-orbit space debris reduction technologies. This satellite demonstrates China's capability for proximity operations, docking, and maneuvering, similar to Northrop Grumman's Mission Extension Vehicle (MEV) satellites [2]. However, unlike the MEV docking with Intelsat, the docking object BeiDou-G2 was a defunct satellite. It was necessary to determine its rotation state in advance to plan suitable strategies and operations for safety.

For the 10-year range of 2010–2019, we accumulated 2964 photometric tracklets, from which 922 reliable rotational frequencies could be extracted (see Extended Data

[1]Purple Mountain Observatory, Chinese Academy of Sciences, Nanjing 210023, China. [2]Key Laboratory of Space Object and Debris Observation, Chinese Academy of Sciences, Nanjing 210023, China. *E-mail: linhouyuan@pmo.ac.cn

Figs. 1 and 2). The data were obtained from the debris surveillance network of Purple Mountain Observatory (PMO), which is the largest dedicated civilian optical network in China and has more than 20 telescope systems across the country. The Graz SLR station provided 15 additional photometry observations with single-photon counters [3], filling the data gap from May to October 2015, when the satellite's subsatellite point drifted westward and out of China's observation range. To ensure the success of Shijian-21's docking task, we needed to limit the error about the rotational state to be less than 5° in the orientation of the rotational axis and 1°/s in rotation speed. This was a demanding requirement. As a comparison, Envisat, another candidate target of active debris removal [4], had a 20° deviation in the early estimation results of the rotational axis [5,6], and then it was found that its rotational axis might vary within a range of about 30° with uncertain law [7-9]. Our expected accuracy was almost an order of magnitude better. Moreover, our purpose was not only to calculate its rotational motion state with these data, but also to accurately predict its future rotational state. Thus, it required our rotational motion model to be accurate enough.

Fig. 1**a** shows the changes in the rotation speed of the BeiDou-G2 satellite in the last 10 years. The rotation speed of the satellite had an increasing trend of oscillation. In the monthly change of the rotation speed of BeiDou-G2 in Fig. 1**b**, the oscillation presents an apparent annual trend. The decelerations appeared from December to May, and the accelerations were from June to November. It was speculated that solar radiation might have been the main factor causing this variation. Hence, we established the solar radiation torque model (see Fig. 2) and performed numerical fitting. In the fitting calculation, the influence of various additional factors that might cause rotation speed changes was evaluated (see Methods and Table 1), for which only gravity-gradient torque and the torque from the central symmetric box needed to be considered. Since we could not find uniform parameters that satisfied the whole 10-year evolution, we divided the data into several segments to fit them separately. Finally, the fittable data segments were named S1–S6 (see Fig. 1 and Extended Data Table 1), where S6 was divided into S6a and S6b because of the long time span, which affected the calculation efficiency. For the other abnormal segments that could not be fitted, six abnormal changes were confirmed and named A1–A6 (see Fig. 1 and Extended Data Table 2). The wrong segmented data could not be fitted down to the noise level, as shown in Extended Data Fig. 3, in which the curve has a significant overall deviation.

The abnormal segments that could not be fitted with numerical models indicated that the BeiDou-G2 satellite did not go through an ordinary life during the past 10 years, which makes the prediction of future rotation speed a big challenge. There were two types of abnormal events other than that of A1. In Fig. 1**a**, the rotation frequency data overlap with the orbital semi-major axis and eccentricity. It is apparent from the figure that A3, A4, and A6 each lasted for approximately 100 days, during which the rotation frequency increased rapidly and the eccentricity dramatically increased, whereas the semi-major axis had almost no abnormal changes. The durations of A2 and A5 were each approximately 1 month, during which the semi-major axis bounced and the eccentricity and the rotation frequency were not significantly abnormal. These anomalous changes that could not be fitted were clearly additional persistent dynamic effects, of which fuel leakage was one of the most likely factors. The two different types of changes, which we called "Continuous Type I" and "Continuous Type II" (see Extended Data Table 2), were suspected to be from different factors, such as leaks from different locations or different types of leakages such as fuel and air bags.

However, the change in A1 was different from these types of changes. The eccentricity changed suddenly at A1. Although the rotation-speed data at that time were missing, the rotational states between S1 and S2 were different (see Fig. 1**c**). This change might theoretically have been a controlled maneuver. However, there was no qualitative change in orbit and attitude. Additionally, it had been 2 years since its last maneuver (officially announced as a failure), and there were no similar incidents before or after. The most important fact was that the satellite was over Africa at the time of A1, which was beyond the visible range of China, and there was no reason to choose an active control at that time. Hence, we suspected that A1 was due to a collision.

It could be determined by propagating the orbits of the pair of two-line elements (TLE) on MJD 56128 and 56132 that the collision time was at 56128D + 12.05 h when the closest distance was 1.5 km, which was within the error range of the GEO target. Additionally, the change of velocity was $\Delta \boldsymbol{v}$ = (0.764, −0.628, 0.171) m/s. This collision could be regarded as an elastic collision [10] because the main components of the satellite were rigid bodies. Because the BeiDou-G2 satellite was in the synchronous orbital belt, it could be assumed that the impact object was in the same orbital plane with a similar semi-major axis. Based on the principles of the conservation of energy and the conservation of momentum, the mass distributions of the impact objects with different eccentricities (different incident directions) and their relative speeds were obtained (see Extended Data Fig. 4). When the eccentricity was equal to 0.005, the mass of the impact object was 2226.3 kg, which was equivalent to that of the BeiDou-G2 satellite. For this case, the impact object might have been another satellite, but the probability of this was extremely low because no other satellite was found in the satellite catalog. When the eccentricity was greater than 0.033, the mass of the impact object was less than 10 kg. Therefore, the impact object was likely to be space debris. All of the collisions with relative speeds less than 1 km/s belonged to the category of slow collisions and might not have caused extensive damage generally [11].

The details of the collision could be further confirmed based on the rotation state. After propagating the fitting results of the rotation states of S1 and S2 (see Extended Data Table 3) to the time of A1, the rotation states before and after the collision could be obtained, as shown in Extended Data Fig. 5, and multiple angular momentum changes could be calculated. The change in angular momentum is generally known to be perpendicular to the impulse (i.e., the same direction as the $\Delta \boldsymbol{v}$). As can be seen from Extended Data Table 4, one angular momentum change was very close to perpendicular to the impulse, whereas the other results deviated significantly. Therefore, it could be determined that this pair of angular momentums formed a valid solution, as shown in the results marked with stars in Extended Data Table 3, and the ecliptic longitude of the solution was determined to be $\lambda_2$. Based on the preceding results, the impact process could be completely described. Fig. 3 shows the results with an eccentricity of 0.2 as an example. In the preceding calculation, we ignored the type II abnormal event that occurred in A2. According to the result of A5, it could be assumed that the type II abnormal event had little effect on the rotation speed. Additionally, we neglected the process in which the angular momentum in the body-fixed coordinate system transferred to the maximum principal axis of inertia because of internal friction. We believed that time scale of this process was short enough because the artificial satellite's fuel and flexible components could easily trigger it.

From Fig. 1**ab**, it is apparent that all abnormal events occurred in the acceleration period (beginning in June to September) and that the events occurred every year after the collision event A1 until 2017. It could be inferred that the collision at A1 was the trigger for the following five abnormal events. A speculation was that the impact was near the fuel tank at the tail. During the annual acceleration period, the Sun illuminated the tail for a long time for one of the symmetric solutions (see Extended Data Figs. 6 and 7). Continued heating changed the material structure of the damaged area, causing fuel leakage. After 2017, no rotational speed abnormalities occurred, probably due to the complete release of the residual fuel. We could expect no similar acceleration events to occur prior to the docking mission.

In addition to A1, special events also occurred in A4 and A6. As shown in Fig. 4**a**, the total reflection coefficient of the front of the solar panels estimated in S4 and S5 decreased abruptly, whereas the ratio of the back-to-front reflection coefficient (B/F ratio) did not change very much. This indicated that the torque on the satellite was weakened, but the overall configuration of the satellite hardly changed. One possibility was that the solar panel was damaged in A4. To verify this conjecture, we developed a model with the solar panel in the +$y$ direction being 25% damaged in S4 and S5 to estimate its front total reflection coefficients. Values that were equivalent to those of S1 to S3 were obtained (the black dots in Fig. 4**a**), as well as the same panel angles and B/F ratio as in the model of the undamaged state. These values showed that our conjecture was reasonable. This damage likely came from the disintegration caused by rapid rotation. The circumstance of the long-term change rate of the rotation speed dropping by an order of magnitude after A4 (see Extended Data Table 1) might also have been related to this. A large gap in the panel angle distributions can also be seen in Fig. 4**b**. Furthermore, at the beginning of A6, on June 29, 2016 (MJD = 57568), National Aeronautics and Space Administration announced that BeiDou-G2 had fragmented into at least five pieces [12] (subsequent reports changed the amount to three pieces), but these fragments had not yet been cataloged in TLE at the time of this study. Fig. 4**a** shows that the B/F ratios of S6a and S6b underwent major changes, indicating that the satellite had indeed experienced structural changes; e.g., its maximum principal axis of inertia might have no longer coincided with the original $z$-axis. In theory, the degree of damage could be estimated by adding undetermined parameters. However, we believe that, as Fig. 1**ef** shows, the existing models for fitting S6a and S6b to each other agreed well with the observations. Any newly added parameter might cause overfitting and yield low-reliability results. In the case where the result of $\lambda_2$ had been confirmed to be the true solution (see Extended Data Table 3), we took the estimation results of S6b as the reference values and propagated the rotational states from S6a to S6b, in which both uncertainties were considered. The standard errors of the rotational axis were 0.5° in the ecliptic longitude and 2.5° in the ecliptic latitude, and that of rotational speed is $3.1\times10^{-4}$ s$^{-1}$, i.e., 0.11°/s. If there were no more abnormal events in the future, the existing parameters and propagating models could satisfy the accuracy requirements of the rotation state at the moment of docking well. Ultimately, Shijian-21 confirmed our predictions.

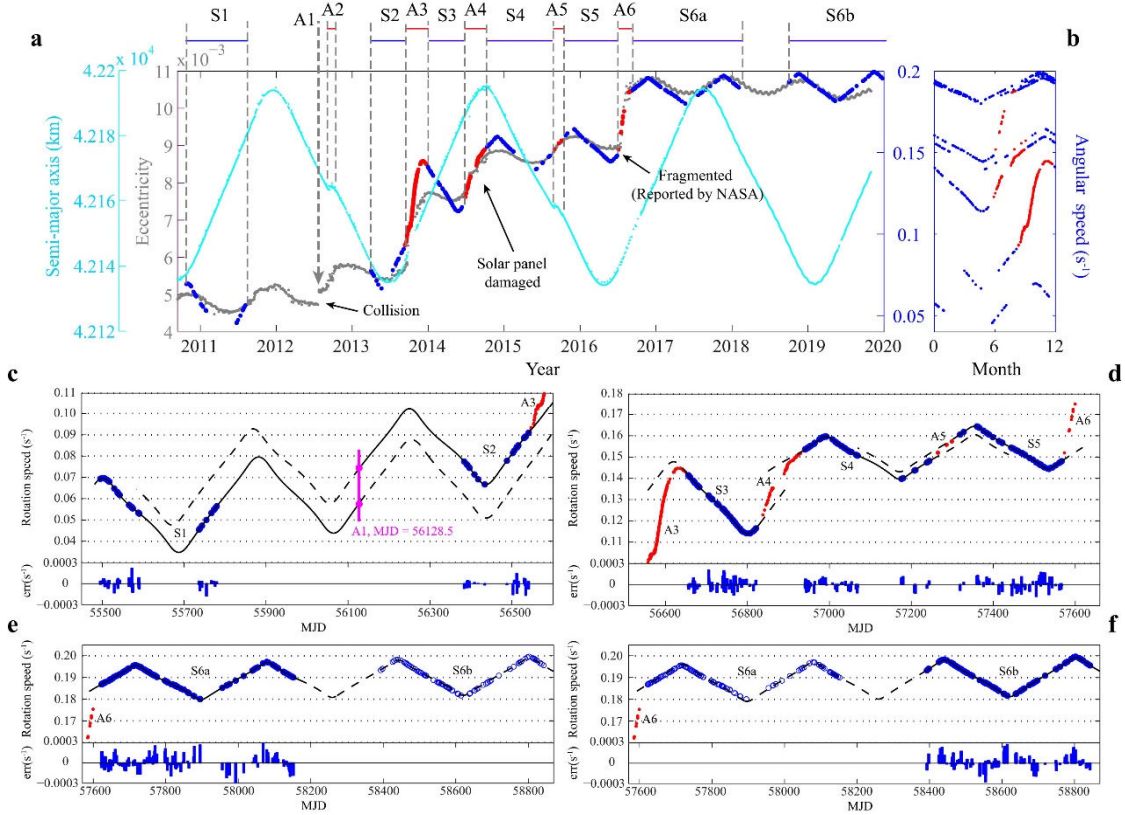

**Fig. 1 | Evolution of the rotation speed of the BeiDou-G2 satellite in the past 10 years and the fitting results. a.** The BeiDou-G2 satellite's rotation speed (blue-red, right ordinate), semi-major axis (green, left, first ordinate), and eccentricity (gray, left, second ordinate). The rotation data were extracted from the photometric data that Purple Mountain Observatory and Graz observed, and the orbit data were derived from the two-line element set. The blue dots, which represent the normal rotational speeds that could be fitted, are divided into seven segments, S1–S6b (see Extended Data Table 1); and the red dots named A1–A6 are in the intervals with abnormal rotational speeds or orbital elements (see Extended Data Table 2). It was inferred that A1 included a collision (see Fig. 3), one solar panel was damaged in A4 (see Fig. 4a), and fragmentation occurred at the beginning of A6. **b**. Monthly changes of the rotation speed of the BeiDou-G2 satellite. The lower four panels show the results of fitting and propagation in **c.** segments S1 and S2; **d.** segments S3, S4, and S5; **e.** segment S6a; and **f.** segment S6b. The bottom of each panel is the fitting residual with weight. The purple in panel **c** marks the time when the fitting results of the rotation states of S1 and S2 were propagated to 56128 days + 12.05 h.

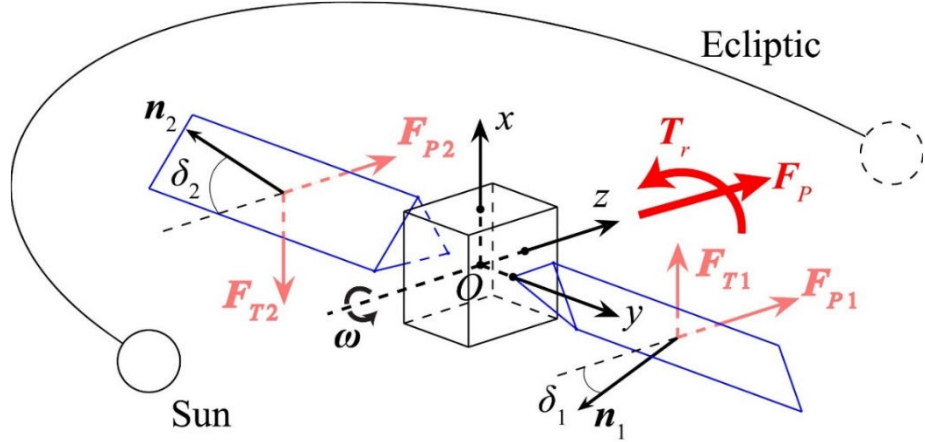

**Fig. 2 | The BeiDou-G2 satellite box-wing model and the solar radiation torque model.** The body-fixed coordinate system *Oxyz* was established along the three principal axes of inertia with the center of mass as the origin. The moments of inertia in the three directions were $I_x$ = 3330.5 kg·m$^2$, $I_y$ = 1717.7kg·m$^2$, and $I_z$ = 3922.8 kg·m$^2$. The black-lined cube in the middle was the main body with the sides' lengths of 2.4 m, 1.72 m, and 2.2 m, along with the *Ox*, *Oy*, and *Oz* directions, respectively. The blue rectangles on both sides were 6 m × 1.68 m solar panels. The vectors $\boldsymbol{n}_1$ and $\boldsymbol{n}_2$ were the normals of the surfaces of the two solar panels with a larger total reflection coefficient (defined as the front), and the angles between them and the +*z* axis were $\delta_1$ and $\delta_2$. When the Sun was in the position shown in the figure, the forces on the solar panels could be decomposed into $\boldsymbol{F}_{T1}$ and $\boldsymbol{F}_{P1}$; and $\boldsymbol{F}_{T2}$ and $\boldsymbol{F}_{P2}$; respectively. Among them, $\boldsymbol{F}_{P1}$ and $\boldsymbol{F}_{P2}$ formed a resultant force $\boldsymbol{F}_P$ (which was not necessarily perpendicular or parallel to any axis), and $\boldsymbol{F}_{T1}$ and $\boldsymbol{F}_{T2}$ formed a torque $\boldsymbol{T}_r$. When $\boldsymbol{T}_r$ and the rotation velocity $\boldsymbol{\omega}$ were in the same sense, the rotation was accelerated. When the Sun moved to the opposite position (the dotted circle in the figure), the torque in the opposite sense of the rotation would slow it down, thereby forming an annual change in the rotation speed.

arXiv:2204.09258

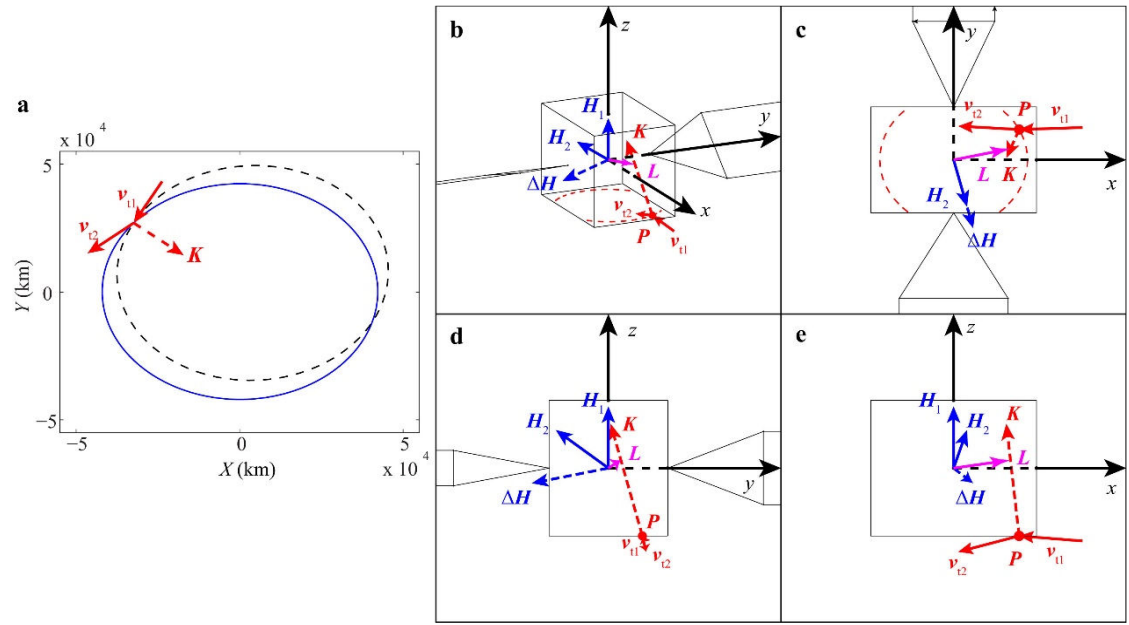

**Fig. 3 | Sketch of the BeiDou-G2 satellite in the collision**. **a**. Sketch of impact orbit. The eccentricity of the impact object's orbit (black dotted line) was assumed as 0.2. **b**. 3-D display of collision in body-fixed coordinate system. (**c, d, e**) Cross-sectional view of impact in 3 coordinate planes. The red solid vectors were the velocity directions of the object before ($v_{t1}$) and after ($v_{t2}$) the collision. The red dotted vector was the direction of the impulse $K$ generated by the collision on the satellite BeiDou-G2. The blue vectors were the angular momentums before ($H_1$) and after ($H_2$) the collision and the change of angular momentum ($\Delta H$). The purple $L$ was the arm of force. The red dashed arcs on the bottom surface of the body were all possible collision points due to the rotation of the satellite.

arXiv:2204.09258

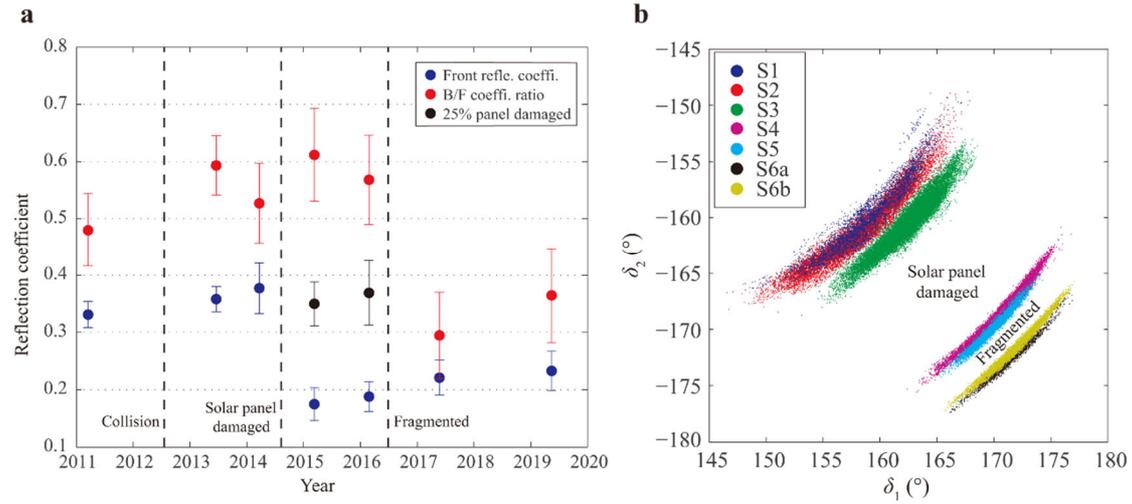

**Fig. 4 | Estimation results of parameters for the BeiDou-G2 satellite's solar panels in all segments. a.** The front total reflection coefficient $r_{sf}+r_{df}$ (blue) and the ratio of back-to-front (B/F) reflection coefficients $(r_{sb}+r_{db})/(r_{sf}+r_{df})$ (red). The black data represented the front total reflection coefficient estimated from the model with 25% of the solar panel in the $+y$ direction being damaged in segments S4 and S5. The estimated reflection coefficient included various factors in the model simulation and was not the actual reflectivity or albedo. (see Methods). **b.** The panel angles in results for $\lambda_2$ (see Methods). The fitting residuals of the results shown in this figure were less than $10^{-4}$ s$^{-1}$ (for S1–S5) and $1.3 \times 10^{-4}$ s$^{-1}$ (for S6a and S6b).

**Table 1.** The order of magnitude of the relative deviation caused by each factor for BeiDou-G2 in its variation of rotational speed from the numerical simulation.

| Factor | Order of magnitude |
| --- | --- |
| Gravity-gradient torque [15] | $10^{-2}$/yr, |
| Provided by the satellite's central symmetric-box (including center of mass and reflection coefficient offsets) | Around $10^{-3}$ |
| Earth's radiation [16] | $10^{-4}$/yr |
| Self-shadow effect | $10^{-5}$/yr |
| Eddy current torque [17] | $10^{-5}$/yr |
| Third-body gravity-gradient torque | $10^{-7}$/yr (Moon) and $10^{-8}$/yr (Sun) |
| Atmospheric density gradient torque [18] | None |

**Methods**

**Dynamical model of the BeiDou-G2 satellite.** The classic box-wing model [13] was adopted to model the satellite's shape, as shown in Fig. 2. The body-fixed coordinate system *Oxyz* was established along the three principal axes of inertia, with the center of mass as the origin *O*. The initial state in which the normal of the front surface of each panel coincided with the +*z* axis was defined, and then, the angles of the two panels rotating around the +*y* axis were $\delta_1$ and $\delta_2$.

The solar radiation torque model was established based on the radiation pressure model proposed by McInnes [14]. Considering the thermal equilibrium, the radiation pressure on the solar panel could be written as

$$\boldsymbol{F} = -\frac{W}{c} A \cos \alpha \cdot \left\{ \left( 2r_s \cos \alpha + B_f \cdot r_d + c_a \frac{\varepsilon_f B_f - \varepsilon_b B_b}{\varepsilon_f + \varepsilon_b} \right) \boldsymbol{n} + (1 - r_s) \cdot \boldsymbol{u} \right\}, \quad (1)$$

where *W* is the flux density, which could be taken as the solar radiation constant $W = 1,361$ W/m$^2$, *c* is the speed of light, *A* is the radiation area, *α* is the angle between the unit normal vector *n* on the surface and the unit vector *u* of the light-source direction, $r_s$ is a specular reflection coefficient, $r_d$ is a diffuse reflection coefficient, $c_a$ is an absorption coefficient, and $r_s + r_d + c_a = 1$. The surface of the solar panel with a larger total reflection coefficient $r_t = r_s + r_d$ was defined as the front. The values $\varepsilon_f$ and $\varepsilon_b$ are the emissivities of the front and back of the panels, respectively. The values $B_f$ and $B_b$ are the non-Lambertian coefficients of the front and back of the panels, respectively, which were both ideally equal to 2/3 in our model. Obviously, the annual oscillation of the rotation speed would be absent if the angles $\delta_1$ and $\delta_2$ of the two solar panels were the same.

We compared the orders of magnitudes of the changes in the rotation speeds of the BeiDou-G2 satellite caused by various factors with the noise of its derived rotation speed, whose relative error was on the order of $10^{-3}$, to evaluate the influence of each factor. The numerical test results are shown in Table 1. Only the gravity-gradient torque and the torque from the central symmetric box needed to be considered. The gravity-gradient torque [15] was inversely proportional to the third power of the distance to the Earth's center, which was expressed as

$$\boldsymbol{M}_G = \frac{3GM}{|\boldsymbol{R}|^5} \boldsymbol{R} \times (\boldsymbol{IR}), \quad (2)$$

where *G* is the gravitational constant, *M* is the mass of the Earth, *R* is the position vector of the center of the Earth relative to the center of mass of the satellite, and *I* is the inertia tensor. For the BeiDou-G2 satellite, numerical verification showed that the calculated *R* from the satellite's mean orbit could already meet the accuracy requirements. The average relative deviation caused by the torque provided by the satellite's central symmetric box was on the order of $10^{-3}$ and had no long-term effect. Therefore, the values of the position of the center of mass and the reflection coefficient of each surface of the box did not need to be very accurate. Additionally, the reflection coefficient of each surface could be approximated by reference to the general materials (see Extended Data Table 5). Ignoring the satellite's self-shadow effect and the Earth's radiation [16] could save a large amount of time in calculation. The eddy current torque, which could be calculated with [17]

$$M_E = B \times (L \cdot (\omega \times B)), \qquad (3)$$

was inversely proportional to the sixth power of the distance to the Earth's center, where $\omega$ is the rotational angular velocity, $B$ is the magnetic field, and $L$ is a defined tensor called eddy current torque tensor. The numerical test, in which the diagonal elements in $L$ were taken to be up to $10^6$ N·m·s/T$^2$, showed that with the effect of the eddy current torque, the relative deviation of the satellite's rotation speed reduction was only about $10^{-5}$/yr, which could be ignored. Other common torques in space could all be ignored, among which the atmospheric density gradient torque [18] that was related to the orbital height decreased sharply as the orbital height increased, and the third-body gravity-gradient perturbations were about $10^{-5}$ with the Moon and $10^{-6}$ with the Sun relative to that of the Earth.

**Selection and estimation of the fitting parameters.** Because the direction of the re-emission force was the same as the direction of the diffuse reflection force, their parameters cannot be estimated separately for the satellite considered in this work. Hence, to reduce the fitting parameters for the surface material characteristics, the thermal emissions were assumed to be identical. Only the reflection coefficients of the front and back surfaces of the solar panels, including the two specular reflection coefficients $r_{sf}$ and $r_{sb}$ and the two diffuse reflection coefficients $r_{df}$ and $r_{db}$, were parameters that needed to be determined. Due to the parameter simplification, the estimated reflection coefficients were not the actual reflectivities or albedos but were rather obtained from the simulation model with the mechanical characteristics of the satellite. They represented the efficiency of the radiation pressure in generating torque, where some deviations between the set values and the actual values, such as the self-irradiation and moments of inertia, were contained.

The extracted frequency and phase-folding values (see Extended Data Figs. 1 and 2) showed that BeiDou-G2 had a good single periodicity, which indicated that the satellite was rotating around the principal axis (not tumbling). This state remained in most observation tracklets and was consistent with the theoretically stable rotational state due to the dissipation effects generated from fuel and flexible components. Hence, in the simulation, we initially set the BeiDou-G2 satellite to rotate around the maximum principal axis of inertia, the *z*-axis. The numerical results showed that the axis of rotation only deviated slightly from the *z*-axis and the result is insensitive to the initial phase angle. Therefore, the only remaining parameters about the attitude to be determined were the initial rotation speed $\omega$, the direction of the rotational axis in space (with the ecliptic longitude $\lambda$ and the latitude $\beta$ being used to facilitate the definition of the direction of the Sun's light).

The remaining parameters to define were the solar panel angles. In the results, there were multiple sets of solutions for the parameters of the solar panels. This was due to the approximate symmetry of the shape model, and the solutions were based on the symmetric solution derived from the definition of the coordinate system. The former could be distinguished through the parameter value-range setting, whereas the latter would be coupled in the solution result. For instance, among all of the parameters, the ecliptic longitude $\lambda$ of the rotation axis was very sensitive to the initial value because it depended on the annual trend of the increase and the decrease of the rotation speed, as shown in Fig. 1**b**, and its rough range could therefore be estimated first. Because of the symmetry of the shape, two sets of values where their ecliptic longitudes were about $\pi$ apart could be obtained in each segment. The parameter $\lambda_1$ was about 130°–160°, and $\lambda_2$ was about 310°–340° (see Extended Data Table 3). Even when the satellite's rotation axis pointed to the same ecliptic longitude value, there were four symmetric solutions for the combination of two solar panels. Hence, for the same rotation state, a total of eight equivalent solutions could be obtained, as shown in Extended Data Figs. 6 and 7. In the two figures, panels **a** and **b** and panels **c** and **d** show the model's rotation around the $+z$ axis or the $-z$ axis, respectively. The difference between the two groups was due to the symmetry of the model. However, the difference between **a** and **b** (or **c** and **d**) was due to the different definitions of the coordinate system, resulting in the different attitudes in the same rotation model at different times. With the assumption that the satellite rotated around the $+z$ axis, the $\delta_1$ and $\delta_2$ solutions could be divided into four zones, as shown in Extended Data Fig. 8, marked as I, II, III, and IV. Each zone had two symmetrical positions corresponding to $\lambda_1$ and $\lambda_2$. The data for the four regions were distributed in a circular shape. According to this, the panel angles $\delta_1$ and $\delta_2$ in the parameters to be determined could be transformed into a radius $r$ from the center of the circle and a phase angle $\varphi$ in simulation. The transformation could be written as the following equations:

$$\begin{aligned} \delta_1 &= \delta_{1c} + r \cdot \cos(\varphi), \\ \delta_2 &= \delta_{2c} + r \cdot \sin(\varphi), \end{aligned} \quad (4)$$

where $\delta_{1c}$ and $\delta_{2c}$ are the angle values of the center of the circle, and the center in each zone of $\lambda_1$ is presented in Extended Data Table 6. From Extended Data Fig. 8, it was easy to obtain the centers of the zone of $\lambda_2$ as $\pi-\delta_{1c}$ and $\pi-\delta_{2c}$. The center value of the equivalent solution rotating around the $-z$ axis differed from these values by $\pi$. The two panels were perpendicular when $r$ was 0, and they were parallel when $r$ increased to $\pi/2$. According to the configuration shown in Fig. 2, the surfaces of the two solar panels irradiated by the Sun were the same in zone II and zone III and were opposite in zone I and zone IV. After the analysis of the rotation evolution in segments S1–S6 (see Extended Data Table 7), all the results showed that most of the fitting results in zone II were the worst, and all the fitting results in zone III were the best. Therefore, the result in zone III could be regarded as the real result, which was also in line with the characteristics of the data, i.e., the acceleration effect was slightly stronger than the deceleration effect. Similar to the reflection coefficients, the panel angle parameters also contained the overall mechanical characteristics of the satellite. According to Fig. 2, an increase in $r$ signified a decrease in the torque generated by the asymmetry. Long-term changes in the panel angle might reveal changes in the mass distribution caused by fuel leakage.

The three degree-of-freedom numerical integration of the Euler kinematic equations and the Euler dynamic equations was adopted in the simulation, and the truncation error accumulated on the order of $10^{-7}$/yr. The nine parameters to be determined, i.e., $r_{sf}$, $r_{sb}$, $r_{df}$, $r_{db}$, $\omega$, $\lambda$, $\beta$, $r$, and $\delta$, were estimated based on the genetic algorithm to fit the satellite's rotation state. The results of the parameters for the solar panels in all segments are shown in Fig. 4, and the other parameters are presented in Extended Data Table 3.

**Data availability**

The data that support the plots within this paper and other findings of this study are available from the corresponding author upon reasonable request.

**Acknowledgements**

This research was supported by the National Natural Science Foundation of China (Grant No. 12073083 and 11533010) and the Youth Innovation Promotion Association, CAS (Grant No. 2018353). The author would like to thank Graz SLR station for providing photometry observation data in 2015. The author thanks for the support and encouragement from DAN&NS, CAS. The author thanks his 4-year-old son Lin Ze-Yu for his company during this work, who was a good kid.

**Author contributions**

H.Y.L. initiated this study, constructed the dynamic model, performed the simulations, analysed the numerical results and drafted the manuscript.

**Competing interests**

The author declares no competing interests.

arXiv:2204.09258

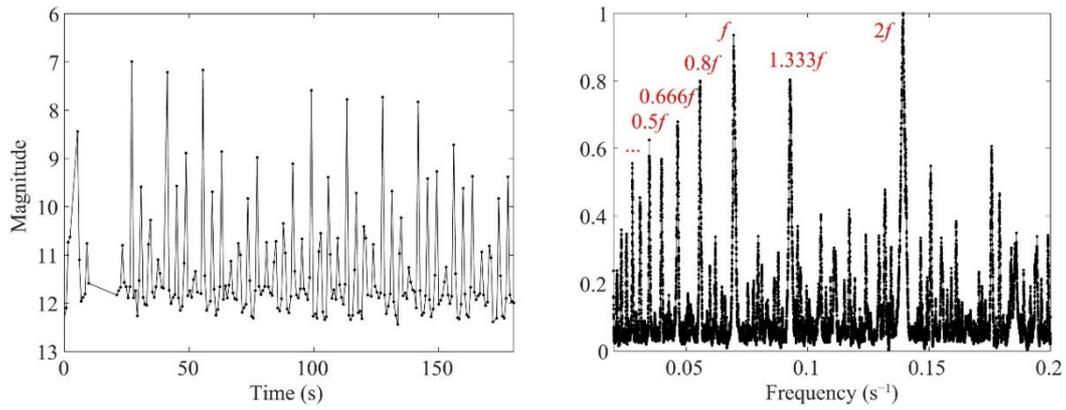

**Extended Data Fig. 1 | Photometry observation data of the BeiDou-G2 satellite on November 4, 2010, and frequency analysis result.** The result contained multiple frequencies with simple integer ratio relationships. From Extended Data Fig. 2, the main frequency could be determined to be $f = 0.06966$ s$^{-1}$.

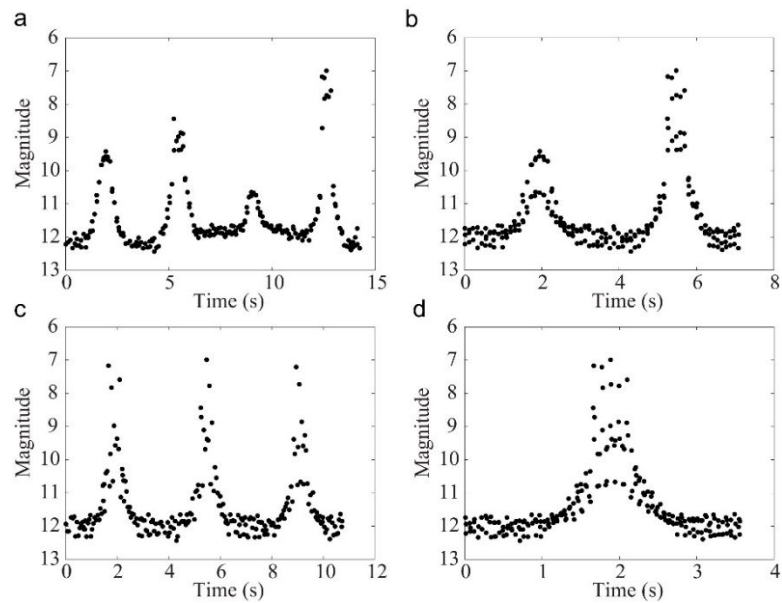

**Extended Data Fig. 2 | Phase folding of observation data on November 4, 2010 (in Extended Data Fig. 1).** Assuming that (**a**) $f = 0.06966$ s$^{-1}$, (**b**) $4f/3$, (**c**) $2f$, and (**d**) $4f$ were frequencies for phase folding, it is easy to observe that taking $f$ as the main frequency is more suitable for the overall curve characteristics. This is also consistent with the characteristic that the cubic structure of the BeiDou-G2 satellite had four brightness peaks (specular reflection).

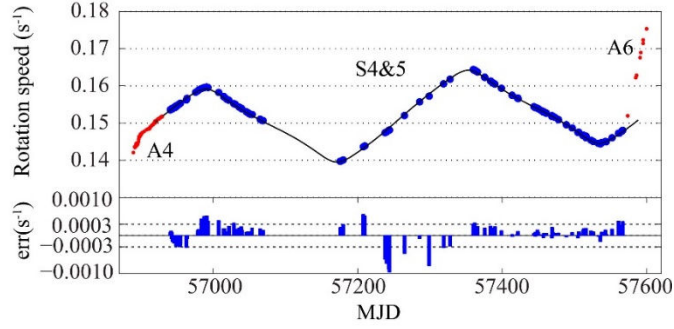

**Extended Data Fig. 3 | The best fit result for a wrong segment with S4, A5, and S5 together.** The fitting residual is $2.51 \times 10^{-4}$ s$^{-1}$.

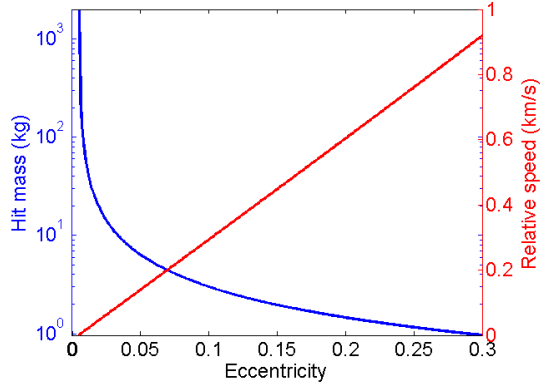

**Extended Data Fig. 4 | Mass and relative speed of the impact object with different eccentricities.** It was assumed that the semi-major axis of the impact object was 42183 km and that its orbit was coplanar with the BeiDou-G2 satellite.

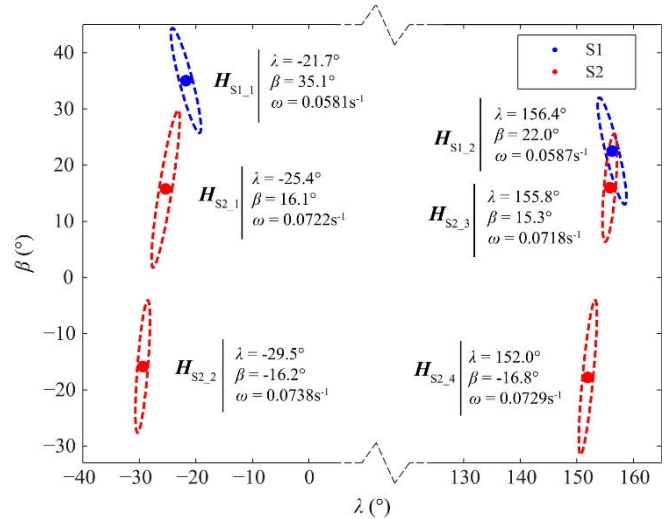

**Extended Data Fig. 5 | Propagation of the angular momentum $H$ of two solutions from S1 (blue) and four solutions from S2 (red) to the time of the collision.** The parameters included the ecliptic longitude $\lambda$, ecliptic latitude $\beta$, and angular speed $\omega$. The filled dot is the average of each solution, and the dashed line marks the three-standard-deviation error ellipse.

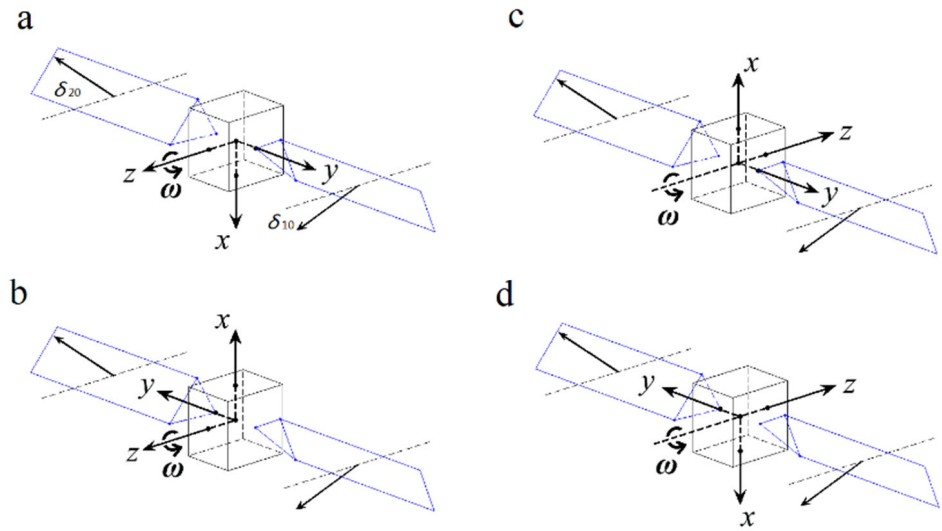

**Extended Data Fig. 6 | The symmetry of the solar panels in the $\lambda_1$ area.** It was assumed that **a**. $\delta_1 = \delta_{10}$, $\delta_2 = -\delta_{20}$. Therefore, **b**. $\delta_1 = \delta_{20}$, $\delta_2 = -\delta_{10}$; **c**. $\delta_1 = \pi + \delta_{10}$, $\delta_2 = \pi - \delta_{20}$; and **d**. $\delta_1 = \pi + \delta_{20}$, $\delta_2 = \pi - \delta_{10}$.

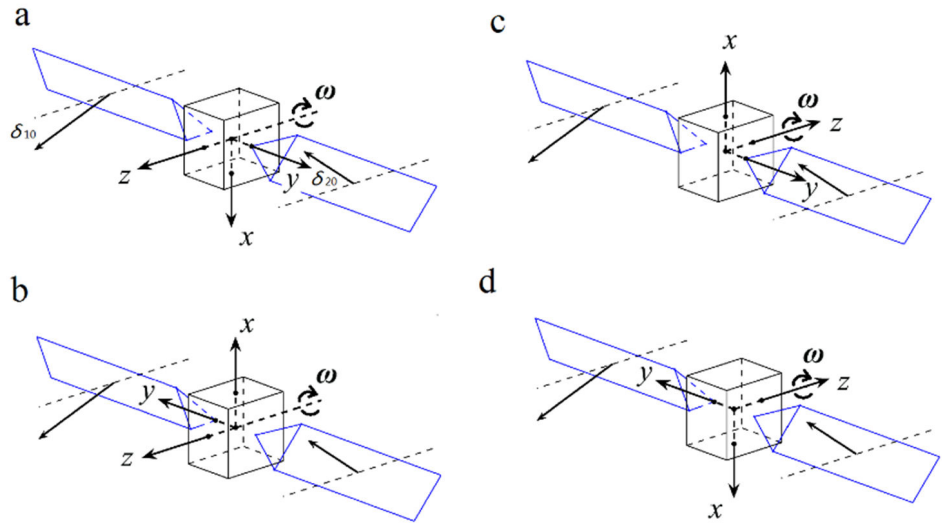

**Extended Data Fig. 7 | The symmetry of the solar panels in the $\lambda_2$ area.** The solution was equivalent to the solutions in Extended Data Fig. 6. According to the definition of Extended Data Fig. 6a, therefore, **(a)** $\delta_1 = -\delta_{20}$, $\delta_2 = \delta_{10}$; **(b)** $\delta_1 = -\delta_{10}$, $\delta_2 = \delta_{20}$; **(c)** $\delta_1 = \pi - \delta_{20}$, $\delta_2 = \pi + \delta_{10}$; and **(d)** $\delta_1 = \pi - \delta_{10}$, $\delta_2 = \pi + \delta_{20}$.

arXiv:2204.09258

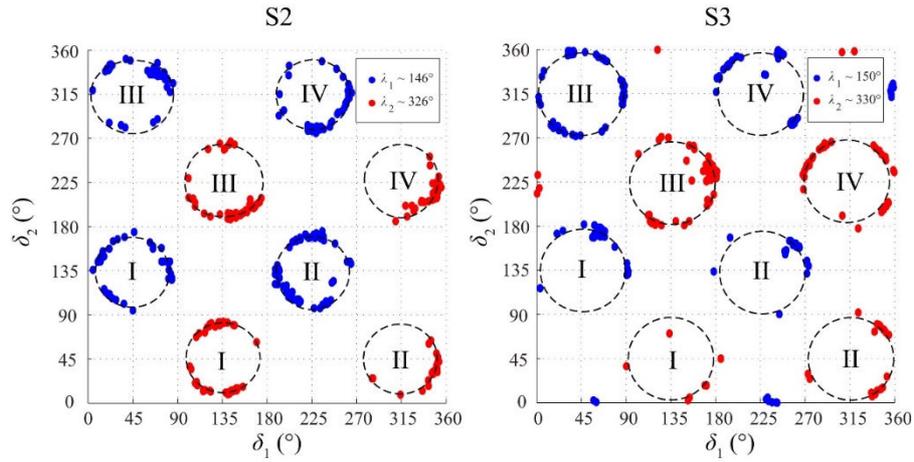

**Extended Data Fig. 8 | Panel angle distributions of the BeiDou-G2 satellite in the preliminary results of rotation around the +$z$ axis in the segments S2 (left) and S3 (right).** The results of the fitting residuals were truncated to less than $10^{-3}$ s$^{-1}$. The blue ($\lambda_1$) and red ($\lambda_2$) colors correspond to the results near different ecliptic longitudes $\lambda$ with symmetrical distributions. The distribution could be divided into four zones by the dashed circles, marked as I, II, III, and IV (see Extended Data Table 6).

arXiv:2204.09258

**Extended Data Table 1** | Change rate of the rotation speed of the BeiDou-G2 satellite in each segment (unit: rad/day$^2$).

| Segment | Short-Term (Observation Data) | | Long-Term (Fitting and Propagating) |
|---|---|---|---|
| | Decreasing | Increasing | |
| S1 | −110 | 139 | 11.8 |
| S2 | −135 | 133 | 12.9 |
| S3 | −102 | / | 12.1 |
| S4 | −67 | 68 | 1.4 |
| S5 | −62 | 68 | 2.1 |
| S6a | −47 | 53 | 1.7 |
| S6b | −50 | 56 | 1.6 |

**Extended Data Table 2** | Abnormal events of the BeiDou-G2 satellite.

| Anomaly | Date | MJD | Change type |
|---|---|---|---|
| A1 | 2012.7.22 (± 2) | 56130 ± 2 | Jump |
| A2 | 2012.9.6–10.5 | 56176–56205 | Continuous type II |
| A3 | 2013.9.10–12.24 | 56545–56650 | Continuous type I |
| A4 | 2014.6.13–10.10 | 56821–56940 | Continuous type I |
| A5 | 2015.8.16–10.15 | 57250–57310 | Continuous type II |
| A6 | 2016.6.29–8.21 | 57568–57621 | Continuous type I |

**Extended Data Table 3 |** Parameters of the rotation axis of BeiDou-G2 at the beginning of each segment. The parameter $\lambda_1$ was about 130°–160°, and $\lambda_2$ was about 310°–340°.

|  | MJD | Ecliptic longitude $\lambda$ (Degree) | Ecliptic latitude $\beta$ (Degree) | Angular speed $\omega$ (s$^{-1}$) |
|---|---|---|---|---|
| S1 | 55495.58282 | 130.1±0.3<br>309.4±0.4* | 34.2±3.1<br>40.0±1.8* | 0.06939±0.00005 |
| S2 | 56383.53552 | 147.8±0.4<br>147.3±0.3<br>327.4±0.4<br>327.0±0.3 * | 16.6±4.5<br>−16.9±5.0<br>16.6±4.5<br>−16.9±5.0 * | 0.07747±0.00004 |
| S3 | 56654.42891 | 148.3±0.4<br>150.7±0.3<br>328.2±0.3<br>330.6±0.4 | −40.8±2.9<br>39.7±2.8<br>−40.6±2.6<br>39.8±2.6 | 0.14177±0.00004 |
| S4 | 56939.01185 | 158.9±0.3<br>160.5±0.3<br>338.8±0.3<br>340.5±0.3 | −41.4±2.3<br>46.3±2.6<br>−40.5±2.6<br>45.5±2.3 | 0.15331±0.00002 |
| S5 | 57318.15202 | 157.0±0.2<br>337.1±0.2 | 51.3±1.8<br>50.6±1.8 | 0.16044±0.00004 |
| S6a | 57637.87944 | 156.1±0.3<br>336.2±0.3 | 51.3±2.0<br>51.6±1.8 | 0.18816±0.00005 |
| S6b | 58392.03400 | 149.3±0.3<br>329.5±0.3 | 56.0±1.8<br>55.1±1.6 | 0.19325±0.00006 |

* The more likely correct solutions in S1 and S2. See Extended Data Fig. 5 and Extended Data Table 4.

**Extended Data Table 4** | Angles between the collision impulse ***K*** and the changes of angular momentums Δ***H*** in different combinations, derived from Extended Data Fig. 5. In theory, the two vectors ***K*** and Δ***H*** should have been perpendicular.

|  | ***H*** $_{s2\_1}$ | ***H*** $_{s2\_2}$ | ***H*** $_{s2\_3}$ | ***H*** $_{s2\_4}$ |
|---|---|---|---|---|
| ***H*** $_{s1\_1}$ | 55.9° | **89.4°** | 160.0° | 172.4° |
| ***H*** $_{s1\_2}$ | 26.7° | 43.2° | 164.8° | 133.0° |

Note: The bold font indicates the best combination of angular momentums that made Δ***H*** closest to perpendicular to ***K***.

**Extended Data Table 5** | Reflectivities on different surfaces of the box body.

| Surface | Surface Material | Reference Specular Reflectivity | Reference Diffuse Reflectivity |
|---|---|---|---|
| ±$x$ | Aluminized polyimide film | 0.77 | 0.1 |
| ±$y$ | Optical solar reflector | 0.95 | 0.03 |
| ±$z$ | Aluminized polyimide film | 0.77 | 0.1 |

**Extended Data Table 6** | Centers of circles of the solar panel angles in each zone of $\lambda_1$ (see Extended Data Fig. 8) when the rotational axis was around the +$z$ axis.

|  | I | II | III | IV |
|---|---|---|---|---|
| $\delta_{1c}$ | 45° | −135° | 45° | −135° |
| $\delta_{2c}$ | 135° | 135° | −45° | −45° |

**Extended Data Table 7** | The residual of the best fit for each zone and segment (Units: ×$10^{-4}$ s$^{-1}$).

| Segment | I | II | III | IV |
|---|---|---|---|---|
| S1 | 2.33 | 6.00 | **0.66** | 2.95 |
| S2 | 0.89 | 1.93 | **0.71** | 0.96 |
| S3 | 0.75 | 0.76 | **0.73** | 0.78 |
| S4 | 0.65 | 0.73 | **0.60** | 0.66 |
| S5 | 0.91 | 0.86 | **0.84** | 0.93 |
| S6a | 1.31 | 2.46 | **1.03** | 1.29 |
| S6b | 1.15 | 2.21 | **1.05** | 1.10 |

Note: The bold font indicates the best fit among all zones.

22